\documentclass[twocolumn,showpacs,superscriptaddress,preprintnumbers,prd]{revtex4}
\usepackage{graphicx,amsmath,dcolumn,bm}
\usepackage{feynmp}
\usepackage[dvipdfm]{pict2e}

\newcommand{\etslash}{E_T \! \! \! \! \! \! \! / \ \ }
\begin{document}
\preprint{KEK-TH-1493, J-PARC-TH-0001}
\title{Test of CDF dijet anomaly within the standard model}
\author{H. Kawamura}
\affiliation{Institute of Particle and Nuclear Studies \\
             High Energy Accelerator Research Organization (KEK) \\
             1-1, Ooho, Tsukuba, Ibaraki, 305-0801, Japan}
\author{S. Kumano}
\affiliation{Institute of Particle and Nuclear Studies \\
             High Energy Accelerator Research Organization (KEK) \\
             1-1, Ooho, Tsukuba, Ibaraki, 305-0801, Japan}
\affiliation{
             J-PARC Branch, KEK Theory Center,
             Institute of Particle and Nuclear Studies, KEK \\
           and
           Theory Group, Particle and Nuclear Physics Division, 
           J-PARC Center \\
           203-1, Shirakata, Tokai, Ibaraki, 319-1106, Japan}
\author{Y. Kurihara}
\affiliation{Institute of Particle and Nuclear Studies \\
             High Energy Accelerator Research Organization (KEK) \\
             1-1, Ooho, Tsukuba, Ibaraki, 305-0801, Japan}
\date{October 27, 2011}
\begin{abstract}
Dijet anomaly reported by the CDF (Collider Detector at Fermilab) 
collaboration in 1.96 TeV $p\bar p$ collisions is investigated
within the standard model by considering effects of parton distribution
functions on various processes: $W$+dijet, $Z$+dijet, $WW$, $ZW$,
and top production. Since the anomalous peak exists 
in the dijet-mass region of 140 GeV with the $p \bar p$ center-of-mass
energy $\sqrt{s}$=1.96 TeV, a relevant momentum fraction $x$ of partons 
is roughly 0.1.
In this $x$ region, recent HERMES semi-inclusive charged-lepton
scattering experiment indicated that the strange-quark distribution
could be very different from a conventional one, which has been 
used for many years, based on opposite-sign dimuon measurements
in neutrino-induced deep inelastic scattering. 
We investigated effects of such variations in the strange-quark 
distribution $s(x)$ on the anomaly. We found that distributions 
of $W$+dijets and other process are affected by the strange-quark 
modifications in wide dijet-mass regions including the 140 GeV one. 
Since the CDF anomaly was observed in the shoulder region
of the dijet-mass distribution, a slight modification of
the distribution shape could explain at least partially the CDF excess.
Therefore, it is important to consider such effects within 
the standard model for judging whether the CDF anomaly indicates 
new physics beyond the standard model. We also show modification
effects of the strange-quark distribution 
in the LHC (Large Hadron Collider) kinematics, where
cross sections are sensitive to a smaller-$x$ region of $s(x)$.
\end{abstract}
\maketitle

\section{Introduction}\label{intro}

A signature beyond the standard model has been searched 
by the Tevatron at Fermilab (Fermi National Accelerator Laboratory)
and by the LHC (Large Hadron Collider) at 
CERN (European Organization for Nuclear Research).
So far, no explicit signature has been found yet.
However, the CDF (Collider Detector at Fermilab) collaboration
reported a possible discovery of new phenomenon in 
high-$p_T$-lepton plus dijet production with missing $E_T$
in $p\bar p$ collisions at 
the center-of-mass energy of 1.96 TeV \cite{cdf-w2j-2011}.
They showed the $W$+dijet cross section and other possible
contributions as a function of the dijet mass $m_{jj}$.
Their data indicated an anomalous peak in the region 
$m_{jj} = 140$ GeV. It is anomalous in the sense
that such an excess is difficult to be explained 
within the standard model.
On the other hand, such a peak cannot be found
in another Tevatron experiment D0 \cite{d0}, so that
it may take time to settle down this issue experimentally.

We may recollect that there was an issue of jet-cross
section excess in a CDF experiment of 1996 \cite{cdf-1996}.
Jet-production cross sections were measured in the $p\bar p$
collisions, and the data were shown as a function of jet transverse
energy ($E_T$). Since the cross sections were significantly larger
than the QCD predications in the energy region $E_T \sim 400$ GeV,
the excess used to be considered as a signature beyond the standard model. 
However, it turned out that the jet excess could be explained if
the gluon distribution is modified at large $x$ ($\sim 0.6$)
\cite{cteq-gluon}, where $x$ is the momentum fraction of a gluon
in the nucleon. 

For explaining the new CDF data of 2011, interesting theoretical
ideas have been proposed \cite{theory-models} mostly by models
beyond the standard model. The ideas include new $Z'$ boson, 
supersymmetry, Higgs boson, relation to string theory, and so on.
However, the experience on the CDF-jet issue of 1996
indicates that careful studies should be done within 
the standard model, particularly by estimating effects 
of parton distribution functions (PDFs).

A fraction of the proton or antiproton beam energy 980 GeV
(=1.96 TeV /2) needs to be transferred to one of the dijets
with the energy about 70 GeV (=140 GeV /2). Therefore,
the relevant region of the momentum fraction $x$ for 
initial partons is $x \sim 0.1$.
This is a very rough estimate because some of the beam
momentum should be carried, for example, by $W$ in the final state.
We need to check whether the PDFs are well determined
in this $x$ region in order to warrant the possibility that
the CDF result is a new discovery beyond the standard model.

The PDFs of the nucleon have been investigated for many years,
and they may be considered to be established except for
extreme kinematical conditions. In fact, the valence-quark
distributions are rather well determined due to abundant 
charged-lepton deep-inelastic-scattering (DIS) data together
with neutrino DIS ones by using strong constraints of charge 
and baryon-number conservations. There is a recent issue of
a nuclear modification difference between the charged-lepton
and neutrino reactions \cite{schienbein}. It affects
the determination of valence-quark distributions
because neutrino-nucleus DIS data are used for extracting
``nucleonic" PDFs. However, it should not be large enough
to affect the CDF anomaly. 

On the other hand, antiquark distributions are not well
determined, particularly for strange and heavy quarks.
The $\bar u$ and $\bar d$ distributions also have some
uncertainties; however, they are relatively well determined
in the region $0.05<x<0.2$ due to Drell-Yan experiments
at Fermilab in combination with DIS measurements \cite{flavor3}.
Furthermore, a new Drell-Yan experiment is currently in progress
at Fermilab \cite{e906},
and it could be continued at other hadron facilities such as
the J-PARC (Japan Proton Accelerator Research Complex)
\cite{j-parc}.
The gluon distribution function $g(x)$ played an important role
in explaining the CDF jet result in 1996 \cite{cteq-gluon}. 
However, it may not be a significant contribution in the present case 
because $g(x)$ is constrained by scaling-violation measurements
of the HERA (Hadron-Electron Ring Accelerator)
except for the distribution in the large-$x$ region.
Therefore, the first parton distribution,
which we need to consider in connection with the new CDF anomaly,
is the strange-quark distribution.

The strange-quark distribution $s(x)$ may have been thought
to be determined from the opposite-sign dimuon measurements 
in neutrino-induced DIS. Current situation is summarized
in Ref. \cite{strange-exp}.
There are some variations among experimental groups; however,
$\int dx x (s+\bar s)$ is 30$-$50\% of the light-antiquark 
distribution $\int dx x (\bar u+\bar d)$. By considering
the neutrino measurements, all the global PDF analysis had been 
done by assuming that $s(x) +\bar s(x)$ has
the same $x$-dependence with $\bar u(x)+\bar d(x)$ until recently. 
There are some recent attempts to assign more flexible functional
forms which are independent of $\bar u(x)+\bar d(x)$
\cite{recent-pdf}. However, the strange-quark distribution
is not accurately constrained as typically shown by large
uncertainties of $s(x)$ by the CT10 parametrization
\cite{recent-pdf}.

In 2008, the HERMES collaboration cast doubt on 
the conventional strange-quark distribution \cite{strange-hermes}. 
They measured semi-inclusive hadron production in charged-lepton DIS 
and they obtained a very different strange-quark distribution
from the ones by global analyses with the neutrino DIS data.
The distribution $s(x)$ is much softer than usual global-analysis 
distributions at $Q^2$=2.5 GeV$^2$. For example, the $s(x)$
is two-times larger than the usual one at $x=0.03$
and it is much smaller at $x=0.2$. Of course, hadron-production
analysis depend much on the employed fragmentation functions,
which have large uncertainties for kaon functions \cite{hkns07}.
Such an ambiguity needs to be considered in evaluating 
the HERMES result seriously.

The light antiquark distributions $\bar u(x)$, $\bar d(x)$, 
and $\bar s(x)$ are considered to be equal in perturbative QCD 
(pQCD) because they are produced from the gluon splitting process 
$g \rightarrow q\bar q$ and their masses are expected to be
negligible in comparison with a hard scale of a reaction.
Of course, higher-order pQCD corrections could give rise to
differences between $\bar u$, $\bar d$, and $\bar s$
because of $Q^2$ evolution with a mixing term $q \leftrightarrow \bar q$
in a splitting function at the next-to-leading order (NLO) together 
with differences between the initial distributions $u$, $d$, 
and $s$ \cite{flavor3}. However, such NLO effects are not 
large enough to explain the measured differences, so that
a non-perturbative mechanism is needed for 
explaining the differences between $\bar u$, $\bar d$, 
and $\bar s$. For the details of these non-perturbative 
models, the reader may look at Ref. \cite{flavor3}.

It is the purpose of this article to discuss whether variations
of $s(x)$, such as the one suggested by the HERMES, could 
affect the CDF anomaly. In particular, cross sections of
$W$+dijet, $Z$+dijet, $WW$, $ZW$, and top production
processes are calculated by using the code 
GR@PPA (GRace At Proton-Proton/Antiproton collisions) 
\cite{gr@ppa} by taking three functional forms of $s(x)$.
The GR@PPA is an event generator for $V$ ($W$ and $Z$)+jets,
$VV$, and QCD multi-jet production processes in $pp$
and $p\bar p$ collisions. 

This article is organized in the following way.
In Sec. \ref{description}, our method is explained for 
calculating the cross sections by the GR@PPA,
and employed strange-quark distributions are explained
in Sec. \ref{strange}.
Results for the lepton-plus-dijet cross sections
are shown in Sec. \ref{results}, and they are
summarized in Sec. \ref{summary}.

\section{Method for calculating cross sections}\label{description}

In describing the cross sections for $W$+dijet, $Z$+dijet, $WW$, $ZW$, 
and top productions, the event generator GR@PPA \cite{gr@ppa} 
is used in this work. 
First, this event generator is explained in Sec. \ref{gr@ppa}. 
We use the CTEQ6L1 distributions for the PDFs in the proton and 
anti-proton. 
Next, in order to find effects of the strange-quark distribution, 
different forms of $s(x)$ are provided in the GR@PPA code
as explained in Sec. \ref{strange}. The other PDFs are not modified
from the CTEQ ones.

\subsection{Event generator GR@PPA}\label{gr@ppa}

The GR@PPA is a framework to calculate cross sections of
proton-proton/antiproton collisions by using the GRACE \cite{grace},
in which Feynman diagrams are automatically computed 
at the one-loop level as well as
the tree level in the standard model.
The GRACE is intended mainly for lepton collisions, so that it is
implemented by including features of hadronic collisions such
as the PDFs in the GR@PPA \cite{gr@ppa}.

The cross section for a certain hadron-hadron reaction process 
can be calculated by the sum of partonic subprocess contributions
\begin{equation}
\sigma = \sum_{i, j, F} \int dx_{1} \, dx_{2} \, d\hat{\Phi}_{F} \,
f^{1}_{i}(x_{1},Q^{2}) \, f^{2}_{j}(x_{2},Q^{2}) \,
{ d\hat{\sigma}_{i j \rightarrow F}(\hat{s}) \over d\hat{\Phi}_{F} },
\label{eq:xsec}
\end{equation}
where $f^{h}_{i}(x_{h},Q^{2})$ is the PDF of the hadron $h$, 
$x_h$ is the momentum fraction carried by the parton $i$ or $j$,
$Q^2$ is a hard scale in the reaction,
and $d\hat{\sigma}_{i j \rightarrow F}(\hat{s})/d\hat{\Phi}_{F}$ is
the cross section for the partonic subprocess $i+j \rightarrow F$.
Here, $i$ and $j$ indicate parton types, $F$ is the final state,
$\hat s$ is the square of the center-of-mass energy in the parton
level ($\hat s=(p_i+p_j)^2$), and $\hat{\Phi}_{F}$ is the phase space
of the final state.

As for the partons, we take into account 11 types
($u$, $d$, $s$, $c$, $b$, $g$, $\bar u$, $\bar d$, $\bar s$, $\bar c$, 
and $\bar b$). There are a huge number of subprocesses which
contribute to the cross section. It is obvious that some of them have
identical diagrammatic structure with slight changes in coupling
constants such as the CKM (Cabibbo-Kobayashi-Masukawa) factors.
Therefore, computation time is significantly reduced by calculating
selected base-subprocesses and then weight factors are simply multiplied:
\begin{equation}
\sigma = \sum_{i,j,F} \int dx_{1} \, dx_{2} \, d\hat{\Phi}_{F} \, 
w_{ijF} \, \frac{d\hat{\sigma}_{i j \rightarrow F}^{\mathrm{selected}}
(\hat{s};m,\alpha)}{d\hat{\Phi}_{F}} .
\label{eqn:cross1}
\end{equation}
Here, $d\hat{\sigma}_{i j \rightarrow F}^{\mathrm{selected}} (\hat{s};m,\alpha_s)$
is the cross section for the base subprocess at the c.m. energy squared,
masses, and couplings. The $w_{ijF}$ is the weight factor which takes into
account the PDFs, CKM matrix $V_{\mathrm{CKM}}$, and decay factors of 
$W/Z$ bosons:
\begin{equation}
w_{ijF} = f^{1}_{i}(x_{1},Q^{2}) \, f^{2}_{j}(x_{2},Q^{2}) \, 
   |V_{\mathrm{CKM}}|^{2K} ,
\label{eqn:cross2}
\end{equation}
where $K$ is the number of $W$ boson in the process.

The GR@PPA code can be obtained from its web page \cite{gr@ppa}.
Outline of GR@PPA  is explained in the following; however,
much details should be found in the original articles \cite{gr@ppa,grace}.
A cross section can be calculated in the following steps.
\begin{itemize}
\vspace{-0.15cm}
\item[(1)] Particles and vertices in theory \\
First, a theoretical model is supplied in the form of
particle content, parameters, and interaction vertices
derived from a model Lagrangian. In this work, the model 
is the standard model of strong and electro-weak interactions.
The model information is supplied as model files.
\vspace{-0.15cm}
\item[(2)] Input for particles and kinematics \\
Initial- and final-state particles as well as reaction kinematics need
to be supplied for calculating cross sections.
The kinematical conditions are c.m. energy, kinematical 
cuts for jets and final-state particles, hard scales 
for the running coupling constant and PDFs, renormalization 
and factorization scales, and the others.
They are supplied in the files grcpar.F and upinit.F
as instructed in Table II of Ref. \cite{gr@ppa}.
The kinematical cuts should be supplied as they were used in
the CDF analysis.
\vspace{-0.15cm}
\item[(3)] Diagram generation \\
From the supplied model information on the particles and vertices,
Feynman diagrams are automatically generated by using mathematical
techniques of graph theory. Here, a node corresponds to a vertex
or an external particle, and an edge does to a propagator or
a connection between a vertex and an external particle.
First, vertices are generated for a given condition of a process.
Then, the vertices are connected by considering all the possible
processes to satisfy obvious conservations such as electric charge
and fermion number. Finally, particles are assigned to the propagators. 
\vspace{-0.15cm}
\item[(4)] Matrix-element generation \\
The generated Feynman diagrams are automatically calculated
by using spinors and gamma matrices in a numerical way.
Based on the Feynman-diagram configuration, the code calls the model
libraries in (1)
for calculating each amplitude.
Here, propagators are expressed by bi-linear forms of wave functions,
Dirac spinors for fermions, and polarization vectors for spin-1 bosons.
\vspace{-0.15cm}
\item[(5)] Monte Carlo integrals \\
Finally, the phase-space integrals of Eq. (\ref{eqn:cross1})
have been calculated by the Monte-Carlo method BASES. 
The parameter NCALL is the number of sampling points 
in integration grids, and it should be large enough to
obtain convergent numerical results. The recommended values
of NCALL are supplied in the subroutine grcpar.F, but they
may be changed according to kinematical conditions.
\end{itemize}

In the published GR@PPA code, the single-top production is not supplied,
so that a separate code is created.
In this work, the single-top and $t \bar t$ production processes 
are calculated, and resulting distributions are shown as top events.

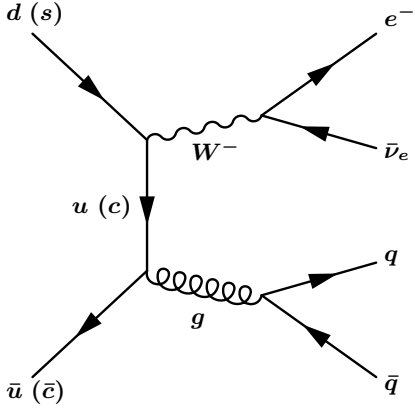
\begin{figure}[h!]
\vspace{0.9cm}
\begin{center}
\begin{fmffile}{./fig-w2j-1}
\begin{fmfchar*}(130,130)
\fmfstraight
  \fmfleft{l1,l2}
  \fmfright{r1,r2,r3,r4}
  \fmf{fermion,tension=1.0}{l2,i2,i1,l1}
  \fmf{photon,tension=1.0}{i2,j2}
  \fmfset{curly_len}{0.25cm}
  \fmf{gluon,tension=1.0}{i1,j1}
  \fmf{fermion,tension=0.5}{r3,j2,r4}
  \fmf{fermion,tension=0.5}{r1,j1,r2}
  \put(-10,-8){\boldmath$\bar u \ (\bar c)$}
  \put(15,62){\boldmath$u \ (c)$}
  \put(-10,135){\boldmath$d \ (s)$}
  \put(60,20){\boldmath$g$}
  \put(60,82){\boldmath$W^-$}
  \put(133,133){\boldmath$e^-$}
  \put(133,83){\boldmath$\bar\nu_e$}
  \put(133,44){\boldmath$q$}
  \put(133,-5){\boldmath$\bar q$}
\end{fmfchar*}
\end{fmffile}
\vspace{0.5cm}
\caption{Typical $W$+dijet process.
The notation $q$ indicates a quark $u$, $d$, $s$, or $c$.}
\label{fig:w2j-1}
\end{center}
\end{figure}

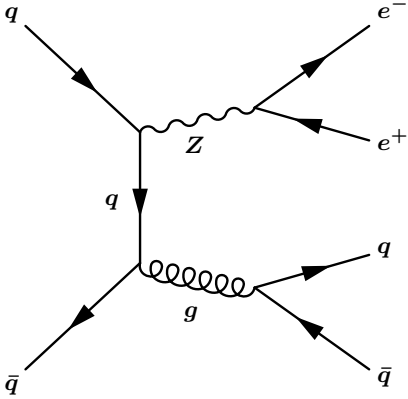
\begin{figure}[h!]
\vspace{0.9cm}
\begin{center}
\begin{fmffile}{./fig-z2j-1}
\begin{fmfchar*}(130,130)
\fmfstraight
  \fmfleft{l1,l2}
  \fmfright{r1,r2,r3,r4}
  \fmf{fermion,tension=1.0}{l2,i2,i1,l1}
  \fmf{photon,tension=1.0}{i2,j2}
  \fmfset{curly_len}{0.25cm}
  \fmf{gluon,tension=1.0}{i1,j1}
  \fmf{fermion,tension=0.5}{r3,j2,r4}
  \fmf{fermion,tension=0.5}{r1,j1,r2}
  \put(-8,-8){\boldmath$\bar q$}
  \put(30,62){\boldmath$q$}
  \put(-8,133){\boldmath$q$}
  \put(60,20){\boldmath$g$}
  \put(60,82){\boldmath$Z$}
  \put(133,133){\boldmath$e^-$}
  \put(133,83){\boldmath$e^+$}
  \put(133,44){\boldmath$q$}
  \put(133,-5){\boldmath$\bar q$}
\end{fmfchar*}
\end{fmffile}
\vspace{0.5cm}
\caption{Typical $Z$+dijet process.}
\label{fig:z2j-1}
\end{center}
\end{figure}

Since there are many processes which contribute to the 
$p+\bar p \rightarrow e^\pm \ (\mu^\pm) + 2 \text{ jets} +\etslash$,
we cannot show all of them here. Only typical processes are
shown. For example, one of $W$+dijet processes is shown
in Fig. \ref{fig:w2j-1}, where $W^-$ is created in an intermediate
stage to produce an electron and an anti-neutrino in the final state,
and two jets are produced by a $q\bar q$ pair from a hard gluon. 
Several more processes are shown in Appendix \ref{appendix-a}
for providing an idea of other contributions. 
For getting information on all the processes, one may look at 
figure files created by running the GR@PPA.
In the same way, typical contributions of
$Z$+dijet, top, $WW$, and $ZW$ processes
are shown in Figs. \ref{fig:z2j-1}, \ref{fig:top-1}, 
\ref{fig:ww0j-1}, and \ref{fig:zw0j-1}, respectively.

In our work, elementary partonic cross sections are calculated
by the GR@PPA. Subsequent parton branching and final fragmentation
into hadrons are not included. They could be calculated, 
for example, by using the event generator PYTHIA \cite{pythia}.
It is the purpose of this work to show gross properties
of strange-quark effects simply by calculating the hard
process part.
In future, if much detailed comparisons become necessary
with the CDF and D0 experimental data,  
they needs to be considered. 
This work is the first step toward such an approach.

\begin{figure}[h!]
\vspace{0.9cm}
\begin{center}
\begin{fmffile}{./fig-top-1}
\begin{fmfchar*}(130,130)
\fmfstraight
  \fmfleft{l1,l2}
  \fmfright{r1,r2,r3,r4,r5,r6}
  \fmf{fermion,tension=0.5}{r1,i1}
  \fmf{fermion,tension=3.0}{i1,l1}
  \fmf{photon,tension=3.0}{i1,i2}
  \fmf{fermion,tension=2.0}{r6,i3,i2}
  \fmf{fermion,tension=1.0}{i2,j1}
  \fmf{fermion,tension=0.3}{j1,r5}
  \fmfset{curly_len}{0.25cm}
  \fmf{gluon,tension=4.0}{l2,i3}
  \fmf{photon,tension=1.0}{j1,j2}
  \fmf{fermion,tension=0.5}{r3,j2,r2}
  \put(-8,133){\boldmath$g$}
  \put(-8,-8){\boldmath$\bar s$}
  \put(15,45){\boldmath$W^+$}
  \put(30,83){\boldmath$b$}
  \put(60,50){\boldmath$t$}
  \put(133,129){\boldmath$\bar b$}
  \put(133,105){\boldmath$b$}
  \put(82,45){\boldmath$W^+$}
  \put(133,50){\boldmath$\mu^+$}
  \put(133,23){\boldmath$\nu_\mu$}
  \put(133,-5){\boldmath$\bar c$}
\end{fmfchar*}
\end{fmffile}
\vspace{0.5cm}
\caption{Typical top-production process.}
\label{fig:top-1}
\end{center}
\end{figure}
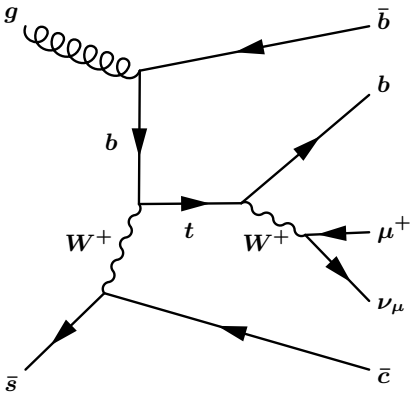

\begin{figure}[h!]
\vspace{0.9cm}
\begin{center}
\begin{fmffile}{./fig-ww0j-1}
\begin{fmfchar*}(130,130)
\fmfstraight
  \fmfleft{l1,l2}
  \fmfright{r1,r2,r3,r4}
  \fmf{fermion,tension=1.0}{l2,i2,i1,l1}
  \fmf{photon,tension=1.0}{i2,j2}
  \fmf{photon,tension=1.0}{i1,j1}
  \fmf{fermion,tension=0.5}{r3,j2,r4}
  \fmf{fermion,tension=0.5}{r1,j1,r2}
  \put(-8,-8){\boldmath$\bar q$}
  \put(30,62){\boldmath$q$}
  \put(-8,133){\boldmath$q$}
  \put(60,22){\boldmath$W^-$}
  \put(60,82){\boldmath$W^+$}
  \put(133,133){\boldmath$\nu_e$}
  \put(133,83){\boldmath$e^+$}
  \put(133,44){\boldmath$d$}
  \put(133,-5){\boldmath$\bar u$}
\end{fmfchar*}
\end{fmffile}
\vspace{0.5cm}
\caption{Typical $WW$ process.}
\label{fig:ww0j-1}
\end{center}
\end{figure}
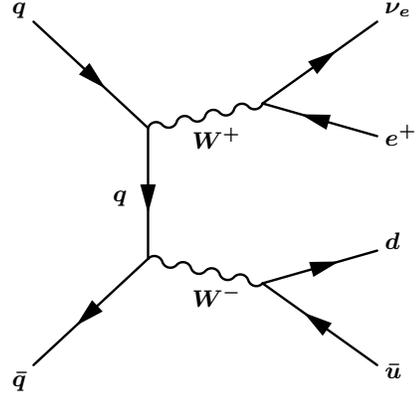

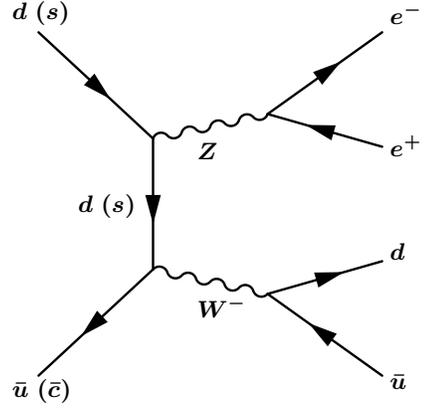
\begin{figure}[h!]
\vspace{0.9cm}
\begin{center}
\begin{fmffile}{./fig-zw0j-1}
\begin{fmfchar*}(130,130)
\fmfstraight
  \fmfleft{l1,l2}
  \fmfright{r1,r2,r3,r4}
  \fmf{fermion,tension=1.0}{l2,i2,i1,l1}
  \fmf{photon,tension=1.0}{i2,j2}
  \fmf{photon,tension=1.0}{i1,j1}
  \fmf{fermion,tension=0.5}{r3,j2,r4}
  \fmf{fermion,tension=0.5}{r1,j1,r2}
  \put(-10,-8){\boldmath$\bar u \ (\bar c)$}
  \put(15,62){\boldmath$d \ (s)$}
  \put(-10,135){\boldmath$d \ (s)$}
  \put(60,22){\boldmath$W^-$}
  \put(60,82){\boldmath$Z$}
  \put(133,133){\boldmath$e^-$}
  \put(133,83){\boldmath$e^+$}
  \put(133,44){\boldmath$d$}
  \put(133,-5){\boldmath$\bar u$}
\end{fmfchar*}
\end{fmffile}
\vspace{0.5cm}
\caption{Typical $ZW$ process.}
\label{fig:zw0j-1}
\end{center}
\end{figure}


\subsection{Applied kinematical conditions in GR@PPA}\label{gr@ppa-change}

In obtaining the cross section, various kinematical cuts 
are applied. They are the same as the ones used 
in the CDF analysis \cite{cdf-w2j-2011}. 
In order to search for the signal 
$p + \bar p \rightarrow WW/ZW \rightarrow \ell \nu_\ell + \text{dijets}$,
events with (two energetic jets) 
+ (one high-$p_T$ electron or muon)
+ (missing $E_T$)
are selected. The high-$p_T$ lepton from $W$ needs to be accompanied
by missing transverse energy ($\etslash$) for a non-interacting
neutrino.

The momentum cut $p_T^\ell >20$ GeV is applied for removing
QCD backgrounds from semi-leptonic decays of hadrons
and $ | \eta^\ell | <1.0$ is due to detector ranges. 
For selecting the leptonic $W$ candidate, a cut is also applied
for the missing $E_T$ as $\etslash >25$ GeV in order to remove 
the QCD backgrounds. Furthermore, the transverse-mass cut 
$M_T^W >30$ GeV also reduces the QCD backgrounds. The cone radius 
$\Delta R = \sqrt {(\Delta\phi)^2+(\Delta\eta)^2}=0.4$
is used for the jet definition with the cuts 
for each jet as $E_T^{\text{jet}} >30$ GeV and $\eta_j <2.4$.
Additional cut conditions are applied as
$| \Delta \eta_{jj} | <2.5$,
$\Delta\phi_{\etslash j_1} >0.4$,
$\Delta R_{j \ell}>0.52$,
and $p_{T \, jj} >40$ GeV
for rejecting back-to-back events,
multi-jet (fake missing $\etslash$) events,
semi-leptonic decays of hadrons,
and for better agreement between data and simulation, respectively.
Gaussian smearing effects are taken into account for $W$ and $Z$ 
by assigning an effective width of 15 GeV in 
the dijet-mass distribution.

\section{Strange-quark distribution function}\label{strange}

\subsection{Definition and experimental information}\label{experiments}

The quark distribution function $q_i (x)$ with flavor $i$ is defined by
the amplitude for the nucleon ($N$) to emit the quark $i$
and then to absorb it at a separated light-cone coordinate $y^-$
\cite{qcd-handbook-1995}
\begin{align}
q_i (x) = \int \frac{dy^-}{4\pi} & e^{-i x p^+ y^-} 
\langle N(p) |  \bar\psi_i(0^+,y^-, \vec 0_{\perp}) 
\nonumber \\
& \times
 \gamma^+ \wp \, \psi_i(0^+,0^-, \vec 0_{\perp}) | N(p) \rangle ,
\label{eqn:def-pdf}
\end{align}
where $\wp$ is the path-ordered ($P$) link to make the expression 
gauge invariant:
\begin{equation}
\wp =  P \exp \left\{  - i g \int_0^{y^-} 
                     dz^- A_a^+ (0^+,z^-, \vec 0_\perp) \, t_a \right\} .
\end{equation}
The $a^\pm$ indicates the light-cone coordinates 
$a^\pm = (a^0 \pm a^3)/\sqrt{2}$,
and $\vec a_\perp$ is the two-dimensional transverse coordinate.
The $\psi$ and $A$ are quark and gluon fields, respectively,
and $t_a$ is the generator of $SU(3)$ group and it is given by
the Gell-Mann matrix $\lambda_a$ as $t_a = \lambda_a/2$.

\begin{figure}[b]
\includegraphics[width=0.37\textwidth]{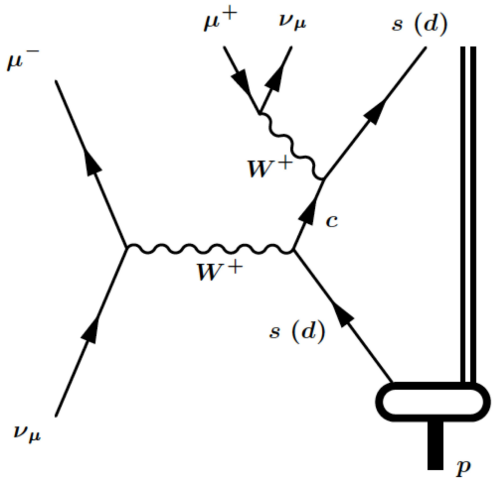}
\vspace{-0.2cm}
\caption{Strange-quark distribution in neutrino-induced 
         opposite-sign dimuon production process
         ($\nu_\mu + p \rightarrow \mu^-  \mu^+ +X$).}
\label{fig:nu-dimuon}
\end{figure}

\begin{figure}[b]
\includegraphics[width=0.37\textwidth]{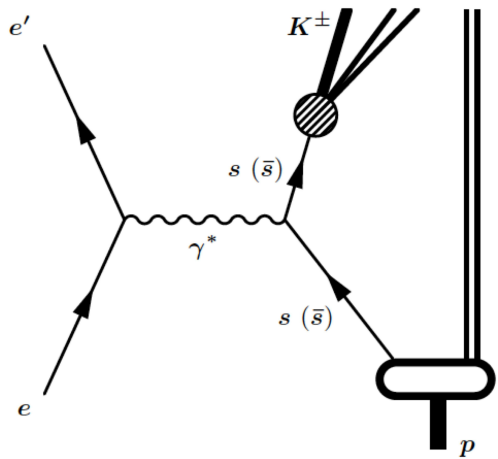}
\vspace{-0.2cm}
\caption{Semi-inclusive kaon production in charged-lepton
         deep inelastic scattering
         ($e + p \rightarrow e' + K +X$).}
\label{fig:semi-inclusive}
\end{figure}

The PDFs can be calculated in nonperturbative theoretical methods 
such as lattice QCD, bag model, chiral soliton model, and so on. 
However, obtained PDFs are not accurate enough to supply precise
cross sections, so that global analyses of world data have been
done for obtaining the optimum PDFs.

As explained in Sec. \ref{intro}, the important PDF to be investigated
is the strange-quark distribution $s(x)$ in connection with
the anomalous CDF result. The strange distribution has been
determined from dimuon measurements in neutrino DIS 
($\nu_\mu + p \rightarrow \mu^-  \mu^+ +X$) \cite{strange-exp}
as shown in Fig. \ref{fig:nu-dimuon}. The dimuon production
occurs from an initial strange (or down) quark 
to produce an intermediate charm quark which decays into
$W^+$ and $s$ (or $d$). From measurements,
the strange-quark-moment ratio to the light-antiquark one 
\begin{equation}
\kappa = \frac{ \int dx \, x \, [s(x,Q^2) + \bar s(x,Q^2)]}
              { \int dx \, x \, [\bar u(x,Q^2) + \bar d(x,Q^2)]} ,
\label{eqn:kappa}
\end{equation}
was obtained. This ratio varies depending on experiments,
but it is in the range of $0.3 \lesssim \kappa \lesssim 0.5$
\cite{strange-exp}.

This neutrino process has been the only constraint for determining 
the strange distribution $s(x)$ in global PDF analyses.
However, the HERMES collaboration recently indicated that 
the distribution $s(x)$ could be much different from the current
one \cite{strange-hermes} by semi-inclusive kaon production
in Fig. \ref{fig:semi-inclusive}. 
The shaded blob in the figure indicates a fragmentation process, which
is described by the fragmentation functions (FFs). It should be noted
that the uncertainties of the FFs are very large for the kaon
as shown in Ref. \cite{hkns07}. For example, such an uncertainty
effect is illustrated in determining the polarized strange-quark
distribution $\Delta s(x)$ from data sets including semi-inclusive 
measurements \cite{lss-strange}.
In any case, considering the large differences between
$s(x)$ of the HERMES and the conventional one,
we think that the strange distribution 
could have much larger uncertainties than the ones suggested
by most PDF analyses \cite{recent-pdf}.

\subsection{Modifications of strange-quark distribution}\label{modifications}

We discuss modifications of the strange-quark distribution
$s(x)+\bar s(x)$ obtained in a typical global analysis.
We define the weight function or modification 
function $w_i (x,Q^2)$ for the strange-quark distribution by
\begin{align}
& [ s(x,Q^2) + \bar s(x,Q^2) ]_{i}
\nonumber \\
& \ \ \ \ \ \ 
= w_i (x,Q^2) \, [s(x,Q^2) + \bar s(x,Q^2)]_{\text{PDF set}} \, ,
\label{eqn:wx}
\end{align}
at any $Q^2$, where perturbative QCD can be applied.
Here, the CTEQ6L1 distributions \cite{cteq6l1}
are used for the PDF set of the global analysis  
because they have been used in the GR@PPA code.
Namely, the function $w_i$ indicates a modification
from the $s+\bar s$ distribution of the CTEQ global analysis. 
The index $i$ indicates a modification type ($i=$1 or 2) explained 
in the following.

\begin{figure}[t!]
\includegraphics[width=0.40\textwidth]{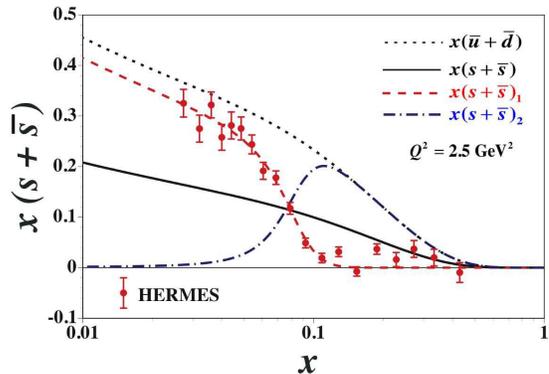}
\vspace{-0.2cm}
\caption{$x(s+\bar s)$ distribution functions are compared with
HERMES data \cite{strange-hermes} and $x(\bar u+\bar d)$. 
The CTEQ6L1 distribution $x(s+\bar s)$ \cite{cteq6l1}
is shown by the solid curve, and the modified distributions
by $w_1 (x,Q_0^2)$ and $w_2 (x,Q_0^2)$ are shown by 
the dashed and dash-dotted curves, respectively.
The weight functions $w_1 (x,Q_0^2)$ and $w_2 (x,Q_0^2)$ 
are explained in the main text. For comparison, 
the $x (\bar u+\bar d)$ distribution is shown 
by the dotted curve.}
\label{fig:ssb-25}
\end{figure}

As the first possible choice, the weight function is determined 
from measurements of the strange-quark distribution 
$s(x)+\bar s(x)$ by the HERMES collaboration.
The average $Q^2$ of the HERMES data is 2.5 GeV$^2$, so that
we define $Q_0^2=2.5$ GeV$^2$ as the scale in providing
the initial PDFs including the modification.
The following hyperbolic-tangent function $w_1$
is assigned for the modification by looking at the differences
between the CTEQ6L1 distribution and the HERMES data 
in Fig. \ref{fig:ssb-25}:
\begin{equation}
w_1 (x,Q_0^2)  = 1 - \tanh \left ( \, \frac{x-x_0}{\Delta x} \, \right ) .
\label{eqn:wx1}
\end{equation}
The HERMES $x(s+\bar s)$ distribution is much larger than 
the CTEQ6L1 at $x<0.06$ and it is much smaller at $x>0.09$.
The parameter $x_0$ indicates the cross-over point, and
$\Delta x$ does the width of the modification curve.
These two parameters are determined so as to fit the HERMES data,
and obtained values are listed in Table \ref{table:parameter-wx}.
The obtained weight function $w_1(x,Q_0^2)$ is shown in 
Fig. \ref{fig:ssb-25}, and it explain the data well except
for slight deviations at $x \sim 0.3$.

\begin{table}[b!]
\caption{Parameter values in the weight functions of
         Eqs. (\ref{eqn:wx1}), (\ref{eqn:wx2}), 
         and (\ref{eqn:wx-q2}) at $Q_0^2$=2.5 GeV$^2$ 
         and $Q^2=2 M_W^2$.}
\label{table:parameter-wx}
\centering
\begin{tabular}{@{\hspace{0.2cm}}c|@{\hspace{0.4cm}}c@{\hspace{0.4cm}}
c@{\hspace{0.4cm}}c@{\hspace{0.4cm}}c@{\hspace{0.2cm}}}
\hline
Function  \      &  $x_0$   &  $\Delta x$   &  $c_0$   &  $c_1$     \\
\hline
$w_1(x,Q_0^2)$ \ &  0.0796  &  0.0253       &  $-$     &  $-$       \\
$w_2(x,Q_0^2)$ \ &  0.0796  &  0.0253       &  $-$     &  $-$       \\
$w_1(x,Q^2)  $ \ &  0.0768  &  0.0408       &  0.802   &  $-$0.475  \\
$w_2(x,Q^2)  $ \ &  0.0727  &  0.0220       &  1.589   &  \ \ 0.417 \\
\hline
\end{tabular}
\end{table}

In addition to the function $w_1$, we consider the following one: 
\begin{align}
w_2 (x,Q_0^2)   = &  \frac{1}{2} \left [ \, 1 
      + \tanh \left ( \, \frac{x-x_0}{\Delta x} \, \right ) 
            \, \right ]
\nonumber \\
& \times
\, \frac{[\bar u(x,Q_0^2)+\bar d(x,Q_0^2)]_{\text{PDF set}}}
        {[s(x,Q_0^2) + \bar s(x,Q_0^2)]_{\text{PDF set}}} \, .
\label{eqn:wx2}
\end{align}
This function $w_2 (x)$ has the opposite functional form
to $w_1 (x)$ in the sense that the small-$x$ part is suppressed
and the large-$x$ distribution is enhanced to reach to
the $\bar u+\bar d$ distribution.
This functions $w_2 \cdot x(s+\bar s)$ is shown 
in Fig. \ref{fig:ssb-25} in comparison with the HERMES data,
$w_1 \cdot x(s+\bar s)$, and $x (\bar u+\bar d)$.
Although $w_2 (x,Q_0^2)$ is not supported by the HERMES data, 
we simply consider it as a trial function to investigate 
sensitivity of the CDF result on the strange distribution.
The strange distribution is roughly constrained by 
the neutrino measurements; however, its $x$-dependent functional 
form is not determined. In particular, the large-$x$
region is not explored experimentally. As there is an issue
of a significant intrinsic charm distribution at large $x$
\cite{intrinsic-charm,global-c}, there may be an enhancement
in the strange at large $x$ \cite{strange-large-x}.
From Fig. \ref{fig:ssb-25}, the small-$x$ distribution may seem 
to be extraordinarily small. However, if the function $w_2$ 
is evolved to the average $Q^2$ ($\sim$20 GeV$^2$) of CCFR 
and NuTeV measurements, the suppression is not very
large as it becomes clear in Sec. \ref{q2-evolution}.
It should be also noted that these functions $w_1$ and $w_2$ 
roughly correspond to the bounds of the $s(x)$ uncertainty
in the CT10 distribution \cite{recent-pdf}.

\subsection{$Q^2$ evolution}\label{q2-evolution}

\begin{figure}[t!]
\includegraphics[width=0.40\textwidth]{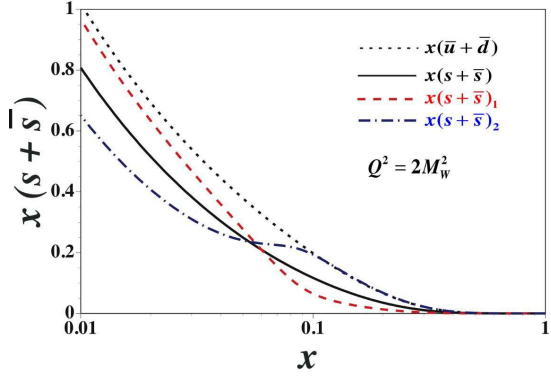}
\vspace{-0.2cm}
\caption{$Q^2$ evolved $x(s+\bar s)$ distribution functions
are shown. 
The CTEQ6L1 distribution \cite{cteq6l1} and the modified 
distributions by $w_1 (x,Q_0^2)$ and $w_2 (x,Q_0^2)$ are 
evolved to $Q^2=2 M_W^2$. For comparison, the $x (\bar u+\bar d)$ 
distribution is shown at $Q^2=2M_W^2$ by the dotted curve.}
\label{fig:ssb-q2}
\end{figure}

The hard scale $Q^2$ in providing the PDFs for the CDF measurements
is much larger than $Q_0^2=2.5$ GeV$^2$, where the HERMES data are
provided for the $s+\bar s$ distribution, so that the strange-quark
distribution needs to be evolved.
As a typical hard scale in calculating the CDF dijet cross sections,
we take $Q^2 = 2 M_W^2$ by considering the dijet mass
$m_{jj} \sim 100$ GeV and the boson (W and Z) masses.
The standard DGLAP (Dokshitzer-Gribov-Lipatov-Altarelli-Parisi)
equations are used for the $Q^2$ evolution by using the code
of Ref. \cite{Q2-code}. In the case without the modification
($w=1$), we checked that the evolved distribution is numerically
consistent with the CTEQ one. The evolved strange distributions 
are shown in Fig. \ref{fig:ssb-q2}.
The strange quarks are copiously produced especially at small $x$
due to $Q^2$ evolution, so that the distributions become large 
at small $x$. Therefore, the differences between
$(s+\bar s)_{\text{PDF set}}$, $(s+\bar s)_1$, and $(s+\bar s)_2$
become smaller than the original ones at $Q_0^2$.
In comparison, the large-$x$ ($>0.1$) distributions are
not so significantly modified by the $Q^2$ evolution. 

\begin{figure}[b!]
\includegraphics[width=0.40\textwidth]{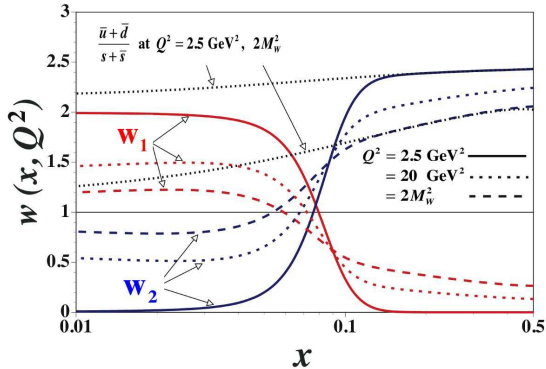}
\vspace{-0.2cm}
\caption{The weight functions $w_1 (x,Q^2)$ and $w_2(x,Q^2)$
are shown at $Q^2$=2.5 GeV$^2$, 20 GeV$^2$, and $2 M_W^2$
by the solid, dotted, and dashed curves, respectively.
For comparison, the ratios $(s+\bar s)/(\bar u+\bar d)$ are
shown at $Q^2$=2.5 GeV$^2$ and $2 M_W^2$.}
\label{fig:wx-q2}
\end{figure}

In order to clearly show the evolution of the weight functions,
we divide the evolved strange distributions $s_i (x,Q^2)$ 
by the CTEQL1 distribution to calculate the weight 
function $w_i (x,Q^2)$ as given in Eq. (\ref{eqn:wx}) at any $Q^2$.
In Fig. \ref{fig:wx-q2}, three $Q^2$ values are taken for showing 
the $Q^2$ evolution of the weight functions $w_1$ and $w_2$. 
They are $Q^2$=2.5 GeV$^2$, 20 GeV$^2$, and $2M_W^2$ 
for the initial HERMES scale, NuTeV scale, and CDF-jet scale,
respectively. Although the modifications are large at $Q^2$=2.5 GeV$^2$ 
and it is in the range $0< w_{1,2} \lesssim 2$,
they become smaller as $Q^2$ increases. 
At the neutrino DIS scale of $Q^2=20$ GeV$^2$, 
the modifications becomes smaller ($0.5 \lesssim w_{1,2} \lesssim 1.5$)
at small $x$ ($< 0.08$). In comparison with $\bar u+\bar d$
at $Q^2=20$ GeV$^2$, the strange to light-antiquark ratio is
$0.23 \lesssim (s+\bar s) / (\bar u+\bar d) \lesssim 0.68$,
which is not very different from the neutrino measurements
\cite{strange-exp} although one should note that 
the neutrino-experimental ratio of Eq. (\ref{eqn:kappa}) 
is for the second moments.
At $Q^2=2M_W^2$, the modifications become much smaller 
and they are in the range $0.8 \lesssim w_{1,2} \lesssim 1.2$
at small $x$ ($< 0.08$).
The modifications also become smaller with increasing $Q^2$
at large $x$ ($>0.1$); however, $Q^2$ variations are not
as large as the ones at small $x$. 

For including the modifications into the event generator,
the evolved weight functions are fitted by the functions
\begin{align}
w_1 (x,Q^2)  = & c_0 
    - c_1 \tanh \left ( \, \frac{x-x_0}{\Delta x} \, \right ) ,
\nonumber \\
w_2 (x,Q^2)   = &  \frac{1}{2} \left [ \, c_0 
      + c_1 \tanh \left ( \, \frac{x-x_0}{\Delta x} \, \right ) 
            \, \right ]
\nonumber \\
& \times
\, \frac{[\bar u(x,Q^2)+\bar d(x,Q^2)]_{\text{PDF set}}}
        {[s(x,Q^2) + \bar s(x,Q^2)]_{\text{PDF set}}} ,
\label{eqn:wx-q2}
\end{align}
where $c_0$, $c_1$, $x_0$, and $\Delta x$ are the parameters
determined by the fitting. Obtained parameters are listed
in Table \ref{table:parameter-wx}. Since it takes time
to calculate the cross sections by the Monte Carlo integrals
in the GR@PPA, these fitted functions are used instead
of calculating the $Q^2$ evolution within the Monte Carlo code.

\subsection{Comments on recent PDFs \\ and theoretical models}
\label{theory}

We comment on the strange-quark distribution in recent global analyses.
Until a few years ago, the distribution $s(x)$ is simply assumed
to be proportional to $[ \bar u (x) + \bar d(x) ]/2$
together with the assumption $s(x) =\bar s(x)$. However,
more flexible parametrizations were recently investigated 
in some analyses by allowing independent $x$-dependence for $s(x)$. 
Then, the uncertainty of the determined $s(x)$ became large
in general.
The size of the uncertainty of $s(x)$ depends on analysis groups. 
A large uncertainty was suggested in the CT10 PDF set 
for the strange-quark distribution as shown in Fig. 12 of 
Ref. \cite{recent-pdf} although uncertainties of other analysis 
groups are smaller.
The CT10 indicates a large uncertainty range $0<w(x,Q^2)<2$
at the scale $Q^2=4$ GeV$^2$ depending on the $x$ region.
This uncertainty range roughly corresponds to the functions
$w_{1,2} (x,Q^2)$ in Fig. \ref{fig:wx-q2}.
Therefore, although the modifications seem to be very large
especially at $Q^2$=2.5 GeV$^2$ in Figs. \ref{fig:ssb-25} 
and \ref{fig:wx-q2}, they are allowed by the current experimental data.
As it is explained in recent global-analysis articles \cite{recent-pdf},
that $Z$ and $W$ production cross sections at LHC are sensitive
to the strange-quark distribution, so that this issue may be solved
in future LHC measurements \cite{sx-lhc}.

Theoretically, the strange-quark distribution stems from
perturbative and non-perturbative mechanisms. The perturbative
creation of $s\bar s$ pairs is described by the 
DGLAP $Q^2$ evolution equations. The non-perturbative contributions
are calculated by phenomenological models such as meson clouds,
chiral soliton, and so on; however, obtained distributions
inevitably depend on the used model. The lattice QCD method
has been making progress; however, it is not at the stage
to provide the detailed $x$ dependent functional form. 
A few low-moments of the PDFs have been obtained so far
in the lattice QCD \cite{lattice-qcd}. 

As a nonperturbative effect on the PDFs, an intrinsic
charm distribution has been often discussed.
The charm distribution $c(x)$ is usually considered to be created
by a perturbative mechanism by the gluon splitting
$g \rightarrow c\bar c$; however, there could exist
the intrinsic charm in the proton according to 
the light-cone Fock space picture \cite{intrinsic-charm}
\begin{equation}
| \, p > = | \, uud> + \cdot\cdot\cdot + | \, uudc\bar c> 
              + \cdot\cdot\cdot .
\label{eqn:intrinsic-c}
\end{equation}
In addition, a meson-cloud picture also suggests 
the intrinsic charm distribution by 
$p (uud) \rightarrow \bar D^0 (u\bar c) \Lambda_c^+ (udc)$,  
         $p (uud) J/\psi (c\bar c)$ \cite{meson-c}.
In the same way, the nonperturbative strange-quark distribution
could exist by considering the strange-quark content
in the proton as \cite{strange-large-x}
\begin{equation}
| \, p > = | \, uud> + \cdot\cdot\cdot + | \, uuds\bar s> 
              + \cdot\cdot\cdot .
\label{eqn:intrinsic-s}
\end{equation}
It is known that the intrinsic charm distribution appears
as a bump in the large-$x$ region as shown in Figs. 2 and 3 
of Ref. \cite{global-c}. 
In the same way, the strange-quark distribution could
have a distribution in the large-$x$ region, which may be 
related to the CDF anomaly. These nonperturbative studies have been
done for light antiquark ($\bar u$ and $\bar d$)
distributions \cite{udbar,flavor3}
and for the strange distribution \cite{strange-large-x}
as well as for the charm \cite{intrinsic-charm}.
In this article, we do not step into the details 
of these mechanisms and their comparisons with the CDF anomaly. 

In future, it is important to investigate the details of 
the theoretical models and the lattice QCD for the strange-quark
distribution, and then their connection to the CDF anomaly. 
Since there is no established theory for the strange-quark distribution, 
we simply vary the CTEQ strange distribution to the softer one 
of the HERMES collaboration and to the harder one as explained
in Sec. \ref{modifications}.

\section{Lepton-plus-dijet cross sections}\label{results}

\begin{figure}[t]
\includegraphics[width=0.40\textwidth]{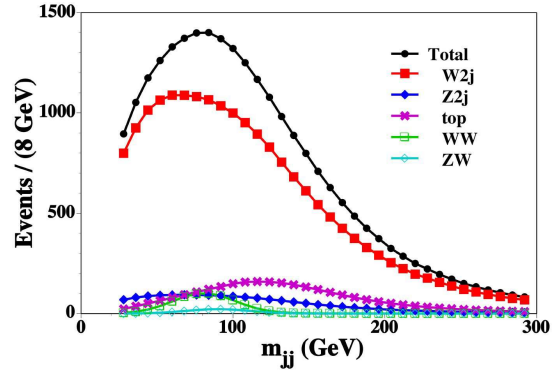}
\vspace{-0.2cm}
\caption{Contributions to lepton-plus-dijet
events in the CDF kinematics.
The $W$+dijet, $Z$+dijet, top, $WW$, and $ZW$
events and their total are shown. Here, the default 
strange-quark distribution is used without modification.
}
\label{fig:wcdf-w2jet}
\end{figure}

Using the GR@PPA code and the modified strange-quark distributions,
we calculate each process contribution to the lepton+dijet
cross section. In particular, $W$+dijet, $Z$+dijet, top, $WW$,
and $ZW$ cross sections are calculated. Using the CTEQ6L1
PDFs without the strange modification, we obtain the events
as a function of the dijet mass in Fig. \ref{fig:wcdf-w2jet}.
We checked that numerical results are consistent with other ones, 
for example, by using the ALPGEN \cite{alpgen}
by calculating these cross sections for Tevatron and LHC kinematics.

The lepton-plus-dijet cross sections are dominated by
the $W$+dijet processes and other contributions are small.
The ordinate indicates the event rate in the dijet mass
interval of 8 GeV. Since the detector acceptance information
is not available for public, and since parton shower and 
final fragmentations are not included in our calculations,
the magnitude of the ordinate cannot be compared with 
the CDF measurements. Therefore, we may
look at overall shapes and each contribution. We find that
the overall features agree with the CDF measurements.

\begin{figure}[b]
\includegraphics[width=0.40\textwidth]{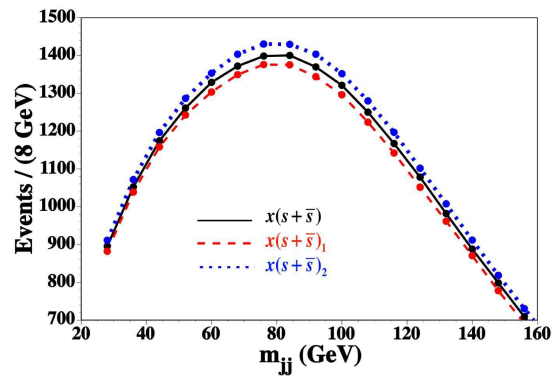}
\vspace{-0.2cm}
\caption{Effects of the strange-quark distribution on
lepton-plus-dijet events in the CDF kinematics.
Three curves are shown by taking the strange-quark 
distribution $s(x)$ without modification and
the distributions multiplied by 
$w_1 (x,Q^2)$ and $w_2 (x,Q^2)$.}
\label{fig:wcdf-w2jet-wx}
\end{figure}

Next, we show the strange-quark effects on the cross section.
The $Q^2$ evolved weight functions $w_1 (x,Q^2)$ and 
$w_2 (x,Q^2)$ are used for calculating the strange-quark
distributions in the GR@PPA. Obtained event rates are shown
in Fig. \ref{fig:wcdf-w2jet-wx}.
It indicates that the cross section increases if the hard
strange-quark distribution $s_2 (x,Q^2)$ is used, and
it decreases if the soft distribution $s_1 (x,Q^2)$ is used.
As mentioned in the introduction section, the $x$ region
which affects the CDF result is $x \sim 0.1$. 
As shown in Fig. \ref{fig:ssb-q2}, the distribution $s(x)$
significantly increases in the hard strange distribution $s_2(x)$
and decreases in the soft distribution $s_1(x)$
at $x \sim 0.1$. This fact gives rise to the modifications
in the lepton+dijet cross sections. 

It is interesting to find that 
the distribution shape as the function of the dijet mass 
becomes wider if the hard strange-quark distribution is used. 
It has a tendency to partially explain the anomalous 
CDF excess, although a sharp CDF-like bump is difficult 
to be obtained within modifications of PDFs. 
However, because the CDF finding is in the shoulder region
of the cross section, a slight shift in the cross-section
shape may explain the CDF excess if they accumulate enough
data to obtain accurate cross sections.

We mentioned a possible bump structure of $s(x)$
in Sec. \ref{theory}. According to our experience on
the numerical estimates, it is not easy to explain 
the sharp CDF peak by a bump in $s(x)$ at medium $x$,
even if it exists, because cross sections are obtained 
by integrating the PDFs over a certain region of $x$, 
more specifically over a certain rapidity range.
In addition, the $Q^2$ evolution smears out the bump.

\begin{figure}[t!]
\includegraphics[width=0.40\textwidth]{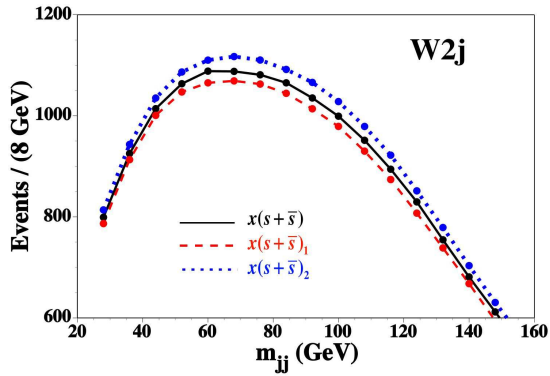}
\vspace{-0.2cm}
\caption{Effects of the strange-quark distribution on
$W$+dijet events in the CDF kinematics.
Three curves are shown by taking the strange-quark 
distribution $s(x)$ without modification and
the distributions multiplied
by $w_1 (x,Q^2)$ and $w_2 (x,Q^2)$.}
\label{fig:w2j-cdf}
\end{figure}

The modification effects are calculated for each process.
We found that there are sizable effects
on the $W$+dijet and $Z$+dijet cross sections from
the strange-quark modification.
However, the effects are very small in top, $WW$, and 
$ZW$ production processes.
Since the magnitude of the cross section is largest
in $W$+dijet, we show the modification effect on this process
in Fig. \ref{fig:w2j-cdf}.
Because the lepton+dijet cross section is dominated
by the $W$+dijet process, the overall modification effects
are similar to the ones in the total event rates in 
Fig. \ref{fig:wcdf-w2jet-wx}. The hard strange-quark distribution
increases the cross section and shifts the shape toward
the larger dijet-mass region, whereas the soft distribution
has opposite effects.

\begin{figure}[t]
\includegraphics[width=0.40\textwidth]{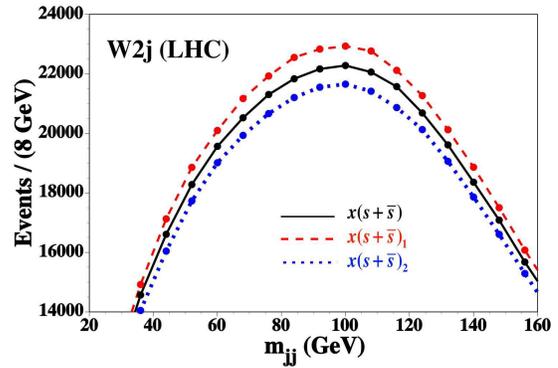}
\vspace{-0.2cm}
\caption{Effects of the strange-quark distribution on
$W$+dijet events in the LHC kinematics.
The notations are the same in Fig. \ref{fig:w2j-cdf}.}
\label{fig:w2j-lhc}
\end{figure}

Next, the same $W$+dijet cross sections are calculated 
for the LHC kinematics of 14 TeV c.m. energy.
Two replacements in the c.m. energy and initial state,
$\sqrt{s}=1.96$ TeV $\rightarrow 14$ TeV
and $p\bar p \rightarrow pp$, are simply done
in the GR@PPA code for the calculation.
The other cut conditions are kept the same simply
for finding kinematical effects.
The obtained results are shown in Fig. \ref{fig:w2j-lhc}.
We notice large differences from the the event rates 
of the CDF kinematics in Fig. \ref{fig:w2j-cdf}.
The cross section decreases (increases) by the hard (soft) 
strange-quark distribution and the shape becomes narrower (wider).
These tendencies are opposite to the CDF ones
in Fig. \ref{fig:w2j-cdf}.
This is simply caused by kinematical effects. Because the c.m. energy
is much larger in the LHC, the process is sensitive to the smaller-$x$
region, particularly at $x \sim 0.02$. Here, the hard distribution
$s_2 (x)$ becomes smaller than the standard one without the modification
at $x \sim 0.02$, and the soft one $s_1(x)$ becomes larger  
as shown in Fig. \ref{fig:ssb-q2}. These differences produce 
the opposite strange-modification effects to
the ones of the CDF kinematics.

These results indicates an interesting conclusion
that the CDF analysis should be carefully done by considering
the uncertainty of the strange-quark distribution.
Furthermore, the strange-quark distribution can be tested
by future LHC experiments. In particular, the extraordinary
HERMES result on $s(x)$ can be tested by both Tevatron and
LHC measurements on vector-boson productions \cite{sx-lhc}.
It is important to note that Tevatron and LHC probe
different $x$ regions of the strange-quark distribution.
There is still a possibility that the CDF anomaly 
could be explained within the standard model
in terms of the PDF modification if the peak 
in the dijet-mass spectrum is not too sharp.

We did not step into the actual calculations based on
some theoretical models. We simply explored effects
due to the modifications of the strange-quark distribution
on the lepton+dijet cross sections. In future, 
it should be interesting to theoretically investigate 
the strange-quark momentum distribution by hadron models 
and possibly by lattice QCD.
We leave them for our future projects.

\section{Summary}\label{summary}

We investigated a possibility that the anomalous CDF result
in the lepton+dijet cross section can be explained within
the standard model by modifications of the parton
distribution functions. 
In particular, the effects of the strange-quark distribution
have been calculated by using the event generator GR@PPA.
The strange-quark distribution $s(x)$ has been thought to be
determined by the opposite-sign dimuon measurements in
neutrino deep-inelastic scattering. However, it became
obvious by the HERMES collaboration that its momentum
distribution, namely the $x$ dependence, has not been
reliably determined at this stage because the much softer
distribution was suggested by their data. 

In this work, two modifications were considered 
for the strange-quark distribution.
One is the soft distribution suggested by the HERMES collaboration,
and another is the hard distribution. The modified distributions
were evolved to the ones at the scale $Q^2 = 2 M_W^2$.
The hard distribution increases the lepton+dijet cross section
and also spread the cross-section shape toward the large dijet-mass
region. It tends to partially explain the CDF excess. 
The sizable strange-quark-modification effects were obtained
for the subprocesses of $W$+dijet and $Z$+dijet productions,
whereas the effects are very small for the subprocesses of 
top, $WW$, and $ZW$ productions.
We found that the PDFs in the region of $x \sim 0.1$ affect
the CDF lepton+dijet cross sections.

The modification effects were also calculated for the LHC
kinematics, and the interesting results were obtained
in the sense that opposite effects are found to the Tevatron case.
The hard distribution decreases the cross section and
shrinks the dijet-mass spectrum, which is opposite
to the ones in the Tevatron kinematics. This is caused
by the fact that the PDFs in the region $x \sim 0.02$
affect the lepton+dijet cross section in the LHC kinematics.
These studies indicate that the $x$ dependence of 
the strange-quark distribution can be investigated
by the $W$ and $Z$ production measurements 
in both Tevatron and LHC. These two-facility experiments
probe different $x$ regions of the strange-quark distribution.

There are still large uncertainties originating
from the strange-quark distribution. If the CDF anomalous
peak is not too sharp, there is still a room to explain
the excess within the standard model in terms of the PDFs
without introducing any new physics beyond the standard model.

\begin{acknowledgements}
\vspace{-0.3cm}
The authors thank 
Y. Miyachi for communications on the HERMES experiment and
S. Odaka for suggestions on the GRAPPA usage.
This work was partially supported by a Grant-in-Aid for Scientific 
Research on Priority Areas ``Elucidation of New Hadrons with a Variety 
of Flavors (E01: 21105006)" from the ministry of Education, Culture, 
Sports, Science and Technology of Japan.  
\end{acknowledgements}

\ \ 

\appendix
\section{An example of $W$+dijet processes}
\label{appendix-a}

\begin{figure}[h]
\includegraphics[width=0.46\textwidth]{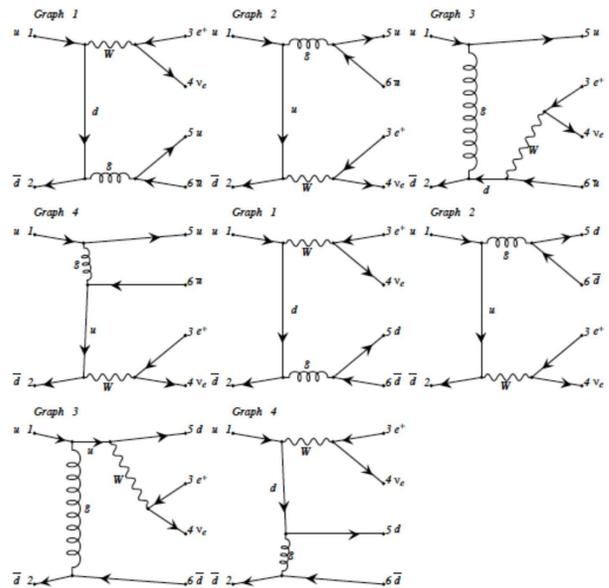}
\vspace{0.0cm}
\caption{Some of $W$+dijet processes are shown by
taking a figure file created by running the GR@PPA code.
These are just a part of all the calculated processes 
for the $W$+dijet events. 
}
\label{fig:w2j-2}
\end{figure}

There are too many processes, which contribute to the $W$+dijet 
cross section, to show all of them.
The figure files to show all the processes are automatically 
generated by running the GR@PPA code.
The reader may look at the processes in postscript
files created by running the GR@PPA.
One of the files is shown in Fig. \ref{fig:w2j-2}.

\vspace{-0.0cm}


\end{document}